\documentstyle[12pt]{article}

\begin{document}
\def\la{\lambda}
\def\debut{\begin{eqnarray}}
\def\fin{\end{eqnarray}}
\def\({\big(}
\def\){\big)}
\def\non{\nonumber}
\def\om{\omega}
\def\proof{{\bf Proof.~~}}
\font\upright=cmu10 scaled\magstep1
\def\stroke{\vrule height8pt width0.4pt depth-0.1pt} 
\def\Cmath{\vcenter{\hbox{\upright\rlap{\rlap{C}\kern
                   3.8pt\stroke}\phantom{C}}}}   
\def\C{\ifmmode\Cmath\else$\Cmath$\fi}

\title {The symplectic structure of rational Lax pair systems.}
 \author{O. Babelon, M. Talon.
 \thanks{L.P.T.H.E. Universit\'e Paris VI (CNRS UMR 7589),
 Bo\^{\i}te 126, Tour 16, $1^{er}$ \'etage,
 4 place Jussieu, F-75252 PARIS CEDEX 05} }
 \date{December 1998}

\begin{titlepage}
\renewcommand{\thepage}{}
\maketitle
\vskip 2cm

\begin{abstract}
We consider dynamical systems associated to Lax pairs depending rationnally
 on a spectral parameter. We show that we can express the
symplectic form in terms of algebro--geometric data provided that the
symplectic structure on $L(\la)$ is of Kirillov type. In particular, in
this case the dynamical system is integrable.
\end{abstract}
\vfill
PAR LPTHE 98-52 \hfill
Work sponsored by CNRS: UMR 7589

\end{titlepage}
\renewcommand{\thepage}{\arabic{page}}

\section{Introduction.}

The algebro--geometrical approaches to integrable systems are the most
powerful ways of solving the Lax equations of motion. These methods
allow to introduce natural coordinates on phase space, namely
$g$ action variables which characterize the spectral curve of genus
$g$ and a divisor of $g$ points on this curve which is equivalent to a
point on the Jacobian and can be related to  angle variables.
These coordinates are also known as the separated Sklyanin's
variables~\cite{Skl}. 
It is an important and natural problem to express the symplectic form
in terms of these coordinates. The first conjecture was made by
Veselov and Novikov in~\cite{NoVe}. More recently the question has
been reconsidered by Krichever and Phong~\cite{KriPhong} who proposed
a general method to prove such a connection. Application of their technique
to the spin Calogero model~\cite{BBKT,BT,Kriche} met some difficulties
and led  to improvements and
simplifications of the method. In particular the role of the branch
points has been recognized in~\cite{Kriche}. The spin Calogero model
however is  rather specific and we propose here to analyze the
situation corresponding to a generic rational Lax pair. In this case
the natural symplectic structure is a coadjoint orbit one, and we find
that we have to use it in all details in order to complete the
calculation of the symplectic form. This shows the nice interplay
between the group theoretical and analytic approaches to integrable
systems. 

\section{The setup.}
Let us consider a $N\times N$ Lax matrix $L(\lambda)$, depending 
 rationally on a spectral parameter $\la \in
\C$, with poles at points $\la_k$
\begin{eqnarray}
L(\lambda) = L_0 + \sum_k L_k \label{laxmatrix}
\end{eqnarray}
where $L_0={\rm Diag}(a_1,\cdots,a_N)$ is independent of $\la$ and
$L_k$ is
the polar part of $L(\la)$ at $\la_k$, ie. $L_k=\sum_{r=-n_k}^{-1}
L_{k,r} (\la-\la_k)^r$.  We assume that $L_k$ lives in a coadjoint
orbit of the group of $N\times N$ matrix regular in the vicinity of
$\la=\la_k$, i.e.
$$L_k = \Big(g_k A_k g_k^{-1}\Big)_-$$
where  $A_k(\la)$ is a diagonal matrix with a pole of order
$n_k$ at $\la=\la_k$, and $g_k$ has a regular expansion at $\la=\la_k$. 
The notation $()_-$ means
taking the singular part at $\la=\la_k$. This singular part only
depends on the singular part $(A_k)_-$ and the first $n_k$
coefficients of the expansion of $g_k$ in powers of $(\la-\la_k)$.  
We see that $(A_k)_-$ specifies the coadjoint orbit, and that the
physical degrees of freedom are contained in the first $n_k$
coefficients of $g_k(\la)$. Since $A_k$ commutes with diagonal
matrices one has to take the quotient by $g_k\to g_k d_k$
where $d_k(\la)$ is a regular diagonal matrix, in order to
correctly describe the dynamical variables on the orbit. 

To interpret $L$ as living in a coadjoint orbit we consider the direct
product of loop groups defined for each $\la_k$. Its Lie algebra is a
direct sum of Lie algebras ${\cal G}_k$ where an element of ${\cal
G}_k$ is a regular sum $X_k(\la)=\sum_{n\ge 0} X_{k,n}(\la-\la_k)^n$
and the Lie bracket is such that $[X_k(\la),X_l(\la)]=0$ if $k\neq
l$. The dual of this Lie algebra can be viewed as $N\times N$ matrix
valued meromorphic functions $L(\la)$ with poles at the $\la_k$ and vanishing
at $\infty$. The duality is expressed by:
$$<L(\la),\oplus_k X_k(\la)>=\sum_k {\rm Res}_{\la_k}{\rm Tr}(L(\la)
X_k(\la))$$ 
Note that $L$ can be uniquely decomposed as $\sum_k L_k$, where $L_k$
is the polar part at $\la_k$ and that only $L_k$ appears in the
residue ${\rm Res}_{\la_k}$. One can add an invariant character $L_0$
to $L$.

The dimension of the orbit of $L_k$ is $N(N-1)n_k$ so that
$L(\la)$ depends on $\sum_k N(N-1)n_k$ degrees of freedom. On this
phase space we have a standard Kirillov symplectic form:
\begin{equation}
\omega =
 \sum_k {\rm Res}_{\lambda_k}\,{\rm Tr}
\((A_k)_- g_k^{-1}\delta g_k \wedge g_k^{-1}\delta g_k\)  \,d \lambda 
\label{Kir}
\end{equation}

In the algebro--geometric approach we study the eigenvector equation:
\begin{eqnarray}
\left(L(\lambda ) - \mu {\bf 1}\right) \Psi(\lambda,\mu ) = 0 
\label{laxfini} 
\end{eqnarray}
and view the eigenvector
$\Psi(\lambda,\mu )$  with eigenvalue $\mu$
as an analytic section of a natural line bundle on the Riemann surface
defined by the characteristic equation:
\begin{eqnarray}
\Gamma~~ :~~ R(\lambda, \mu ) \equiv \det( L(\lambda )-\mu~{\bf 1} )
=0  
\label{specurve} 
\end{eqnarray}
 If $N$ is the dimension of the Lax matrix, the
equation of the curve is of the form:
\begin{eqnarray}
\Gamma ~~ : ~~ R(\lambda, \mu )\equiv (-\mu)^N +\sum_{q=0}^{N-1} 
r_q(\lambda) \mu^q =0 \label{specurve2}
\end{eqnarray}

The coefficients $r_q(\lambda)$ are polynomials in the matrix elements
of $L(\la)$ and therefore have poles at $\la_k$. Since the Lax
equation $\dot L = [L,M]$ is isospectral, these coefficients are
time-independent and are related to the action variables.

From eq.(\ref{specurve2}), we see that the spectral curve appears as
an $N$-sheeted covering of the Riemann sphere. To a given point $\lambda$
on the Riemann sphere there correspond $N$ points on the curve whose
coordinates are $(\lambda, \mu_1), \cdots (\lambda, \mu_N)$ where the
$\mu_i$ are the solutions of the algebraic equation $R(\lambda, \mu
)=0$.  By definition $\mu_i$ are the eigenvalues of $L(\lambda)$.

We assume for simplicity that all the $a_i$'s are different. Then on
the spectral curve, we have $N$ points $Q_i\equiv (\lambda = \infty,
\mu_i = a_i)$ above $\lambda = \infty$. The analyticity properties of
$L(\la)$ are invariant under conjugation by constant matrices. To
preserve the normalization at $\infty$ these matrices have to be diagonal.
 Generically, these
transformations form a group of dimension $N-1$ and we will have to
factor it out.

Before doing complex analysis on $\Gamma$, one has to determine its
genus. A general strategy is as follows. As we have seen,
$\Gamma$ is a $N$-sheeted covering of the Riemann sphere.  There is a
general formula expressing the genus $g$ of an $N$-sheeted covering of a
Riemann surface of genus $g_0$ (in our case $g_0 =0$). It is the
Riemann-Hurwitz formula: 
\begin{eqnarray} 2g-2 = N(2g_0 -2) + \nu
\label{RiemHur}
\end{eqnarray}
where $\nu$ is the branching index of the covering. The branch points
occur at the zeroes of $\partial_\mu R$. The number of its zeroes is
the same as the number of its poles, which are located above the
points $\la_k$. One gets:
$$g={N(N-1)\over 2}\sum_k n_k -N +1$$

For completeness of the method it is important to observe that the
genus is equal to the number of action variables occurring as
independent parameters in the eq.(\ref{specurve2}) which should also
be half the dimension of phase space. This phase space $\cal M$ is the above
coadjoint orbit, quotiented by the action of constant diagonal matrices.
The  orbits of this action are
of dimension $(N-1)$, since the identity does not act. One has to perform a
Hamiltonian reduction by this action. First one fixes the momentum,
yielding $(N-1)$ conditions, and then one takes the quotient by the
stabilizer of the momentum which is here the whole group since it is
Abelian. Hence the dimension of the phase space is reduced by
$2(N-1)$, yielding:
$${\rm dim}\,{\cal M}=(N^2-N)\sum_k n_k -2(N-1)=2g$$

Let us now count the number of independent coefficients in
eq.(\ref{specurve2}). It is clear that $r_j(\la)$ is a rational
function of $\la$. The value of $r_j$ at $\infty$ is known since
$\mu_j\to a_j$. Moreover $r_j(\la)$ has a pole of order $jn_k$ at
$\la=\la_k$. Hence it can be expressed on $j\sum_k n_k$ parameters
namely the coefficients of all these poles.  So we have altogether
${1\over 2} N(N+1) \sum_k n_k$ parameters. They are not all independent
however. Indeed above $\la=\la_k$ the various branches can be written:
$$\mu_j=\sum_{n=1}^{n_k} {c^{(j)}_n\over (\la-\la_k)^n}+{\rm
regular}$$ where all the coefficients $c^{(j)}_1,\cdots,c^{(j)}_{n_k}$
are fixed and non--dynamical because they are the matrix elements of
the diagonal matrices $(A_k)_-$, while the regular part is
dynamical. This implies $Nn_k$ constraints on the coefficients of
$r_j$ (note that $r_j$ is the symmetrical function
$\sigma_j(\mu_1,\cdots,\mu_N)$ hence the $n_k$ highest order terms in
$r_j$ are fixed). We are left with ${1\over 2} N(N-1) \sum_k n_k$ parameters,
that is $g+N-1$ parameters.

It remains to take the quotient by the action of constant diagonal matrices.
Consider the Hamiltonians $H_n=(1/n)\,{\rm res}_{\la=\infty}{\rm Tr}
\,(L^n(\la))\, d\la$,
i.e. the term in $1/\la$ in ${\rm Tr}\,(L^n(\la))$. These are
functions of the $r_j(\la)$. We show that they are the generators of
the diagonal action. First we have:
\begin{eqnarray*}
{\rm Res}_{\la=\infty}{\rm Tr}\,(L^n(\la)) d\la &=& 
n\, {\rm Res}_{\la=\infty}{\rm Tr}\,(L_0^{n-1}\sum_k L_k(\la)) d\la\\ &=&
n\, {\rm Res}_{\la=\infty}{\rm Tr}\,(L_0^{n-1}L(\la)) d\la
\end{eqnarray*}
since all $L_k$ are of order $1/\la$ at $\infty$. 
Computing with the above Kirillov bracket  one obtains
$\{H_n,L(\mu)\}= -[L_0^{n-1},L(\mu)]$ which is the coadjoint action of a diagonal
matrix on $L(\mu)$. Since $L_0$ is generic the $L_0^n$ generate the
space of all diagonal matrices, so we get exactly $N-1$ generators
$H_1,\cdots,H_{N-1}$. In the Hamiltonian reduction procedure, the
$H_n$ are the moments of the group action and are to be set to fixed
(non--dynamical) values.  Hence when the system is properly reduced we
are left with exactly $g$ action variables. 

The eigenvector line--bundle has Chern class $-(g+N-1)$. We get a
non--vanishing section by requiring that the eigenvector $\Psi(P)$ has
its first component equal to 1. Then it has $(g+N-1)$ poles on the
spectral curve~\cite{BBKT}. Since $L$ is diagonal at $\infty$ the
natural eigenvectors of $L(\la)$ above $\la=\infty$ are the canonical
basis vectors. Since we impose that $\Psi_1=1$ this implies that
$\Psi(P)$ has poles at $(N-1)$ points above $\la=\infty$, hence we are
left with $g$ dynamical poles. Note that this is the number of angle
variables and we are led to describe the system in terms of the $g$
action variables entering the equation of the spectral curve and $g$
points on this curve, namely the dynamical poles of $\Psi$.

\section{Symplectic form.}

We have seen that the Lax pair description of a dynamical system
naturally provides coordinates on phase space, namely $g$ independent
action variables $F_i$ which parameterize the spectral curve $\Gamma$, 
and $g$ points $\nu_i=(\la_i,\mu_i)$ 
on the spectral curve, which we called the dynamical divisor.  
It is important to express the
symplectic form in terms of these coordinates. The phase space appears
as a fibered space whose base is the space of moduli of the spectral
curve, explicitly described as coefficients of the equation
$R(\la,\mu)=0$ of the spectral curve, and the fiber at a given $R$ is
the Jacobian of the curve $R=0$. On this space we introduce a
differential $\delta$ which varies the dynamical variables $F_i$,
$\la_i$, $\mu_i$ subjected to the constraint
$R_{\{F_i\}}(\la_i,\mu_i)=0$. 

We will need an auxiliary fiber bundle above the same base 
whose fiber is $\Gamma \times {\rm Jac}(\Gamma)$. We extend $\delta$
to this space by  keeping the previous definition on the ${\rm
Jac}(\Gamma)$ part and on the $\Gamma$ part, we
differentiate any function of $F_i$, $\la$, $\mu$
with $R_{\{F_i\}}(\la,\mu)=0$ keeping $\la$ {\em constant}. 
Remark that $\la$ is  universally defined  on the whole family of curves. 
For a function $f(P;F_i)$ if we take $\la$ as a local
parameter  $\delta f=\sum_i \partial_{F_i} f \delta F_i$. At a
branch point  the local parameter is $\mu$ and
we have:
\begin{equation}
\delta f= \partial_\mu f \delta\mu + \sum_i \partial_{F_i} f \delta
F_i, \quad {\rm with} ~ \delta\mu=-{1\over \partial_\mu R_{\{F_i\}}(\la,\mu)}
\sum_i \partial_{F_i}R_{\{F_i\}}(\la,\mu)\delta F_i
\label{lepiegeacons}
\end{equation}
Note that at a branch point 
$\partial_\mu R_{\{F_i\}}(\la,\mu)=0$, hence the differential $\delta f$
acquires a pole even though $f$ is regular. Remark however that if $f$
depends rationally on $\la$ and the $F_i$, $\delta f$ is regular
at the branch points.

At each point $P(\la,\mu)$ on $\Gamma$ is defined a column eigenvector
$\Psi(P)$ of the Lax matrix up to normalization. This allows to define
a matrix $\widehat{\Psi}(\la)$ whose columns are the $N$ vectors 
$\Psi(P_i)$ at the $N$ points $P_i$ above $\la$ (assuming we have
chosen locally some ordering of the sheets). We also consider its
inverse matrix $\widehat{\Psi}^{-1}(\la)$ and denote its $N$ lines
by $\Psi^{(-1)}(P_i)$ (for the same ordering of the $P_i$).
In particular we have $<\Psi^{(-1)}(P),\Psi(P)>=1$. Note that
$\Psi^{(-1)}(P)$ has poles at the branching points of the covering
$(\la,\mu)\to \la$ since the determinant of $\widehat{\Psi}$ vanishes
here.

We define a three--form $K$ on our extended fiber bundle, and
we regard it as a one-form on $\Gamma$ whose coefficients are
two--forms on phase space.
\begin{eqnarray}
K &=& K_1 + K_2 + K_3 \label{KriPhong} \\
K_1 &=& <\Psi^{(-1)}(P)\delta L(\la) \wedge \delta \Psi(P)>\, d\la \non\\
K_2 &=& <\Psi^{(-1)}(P)\delta \mu \wedge \delta \Psi(P)>\, d\la \non\\
K_3 &=& \delta \( \log \partial_\mu R\) \wedge \delta \mu \, d\la \non
\end{eqnarray}

\proclaim Proposition.
Let us define the two--form on phase space:
$\omega=\sum_{k,i}\, {\rm Res}_{P_{k,i}} \,K $ 
where $P_{k,i}$ are the points above the poles $\la_k$ of $L(\la)$.
Then we have:
$$\omega=2 \sum_{i=1}^g \delta \la_i \wedge \delta\mu_i$$
hence $\omega$ is a symplectic form on phase space.
\par
\proof The sum of the residues of $K$ seen as a form on $\Gamma$
vanishes. The poles of $K$ are located at four different places, first
the dynamical poles of $\Psi$, then the poles at the $P_{k,i}$ coming
from $L$ and $\mu$, next the poles above $\la=\infty$ coming from
$\Psi$ and $d\la$, and finally the poles at the branch points of the
covering coming from the poles of $\Psi^{(-1)}$ and from
eq.(\ref{lepiegeacons}). 

Let us compute the residues at the  dynamical poles
$(\nu_1,\cdots,\nu_g)$. We write the coordinates of these points as:
$\nu_i=(\lambda_i, \mu_i)$ for $i =1,\cdots g$. Near such a point we
can choose $\la$ as a universal local parameter and
$\Psi= 1/(\la -\la_i)\times \Psi_{\rm reg}$ hence:
\begin{eqnarray}
\delta \Psi
= {\delta \lambda_i \over \lambda -
\lambda_i }\left( \Psi + O(1)\right) \nonumber
\end{eqnarray}
Since $(L-\mu)\Psi=0$ and $\Psi^{(-1)}(L-\mu)=0$, we have $(\delta L - \delta \mu)\Psi
+(L-\mu)\delta \Psi=0$. Multiplying by $\Psi^{(-1)}$ we get
$\Psi^{(-1)} \delta L \Psi=\delta \mu$, therefore:
$${\rm Res}_{\nu_i} K_1 =  \delta \mu\vert_{\nu_i}\wedge \delta\lambda_i$$ 
Here $\delta \mu$ is to be seen as a meromorphic function on $\Gamma$
given by eq.(\ref{lepiegeacons}). However varying
$R(\la_i,\mu_i)=0$ we obtain:
$$\delta \mu\vert_{\nu_i}=\delta \mu_i+\left. {\partial_\la R \over
\partial_\mu R}\right\vert_{\nu_i} \delta \la_i$$
and the second term does not contribute to the wedge product.

The contribution of $K_2$ is exactly the same. Finally, $K_3$ is regular
at $\nu_i$ and does not contribute to the residue at this point.
So we finally get:
\begin{equation}
{\rm Res}_{\nu_i} K = 2 \delta \mu_i \wedge \delta\lambda_i
\label{novikov}
\end{equation}

We now show that there are no residues at the branch points due to the 
proper choice of $K_2$ and $K_3$. Let us
look at the term $K_1$.
At a branch point $b$, $\Psi^{(-1)}$ has a simple pole, $\delta L$ is
regular, $\delta \Psi$ has a simple pole due to
eq.(\ref{lepiegeacons}) and the form $d\la$ has a simple zero, hence
the considered expression has a simple pole at $b$. To compute its
residue it is enough to keep the polar part in  $\delta \Psi$, i.e. to
replace $\delta \Psi$ by $\partial_\mu \Psi \delta \mu$ (recall that
$\mu$ is a good local parameter around $b$). We get:
\begin{eqnarray*}
{\rm Res}_b K_1 &=&
{\rm Res}_b <\Psi^{(-1)}\delta L  \partial_\mu \Psi>\wedge \delta \mu \,
d\la \\
&=& {\rm Res}_b <\Psi^{(-1)}(\delta L-\delta\mu)  \partial_\mu
\Psi>\wedge \delta 
\mu \,d\la
\end{eqnarray*}
where in the last equation we have used the antisymmetry of the wedge
product to replace $\delta L$ by $\delta L-\delta\mu$. 
Using again the eigenvector equation $(L-\mu)\Psi=0$, and varying the
point $(\la,\mu)$ on the curve around $b$ one gets
\begin{equation}
(L-\mu)\partial_\mu \Psi= \Psi - {d\la\over d\mu}{dL\over d\la}\Psi
\label{variation}
\end{equation}
 where $d\la/d\mu$
vanishes at the branch point. We then differentiate with $\delta$ and
get:
\begin{eqnarray*}
{\rm Res}_b <\Psi^{(-1)}(\delta L- \delta \mu)  \partial_\mu \Psi>\wedge
\delta \mu\, d\la
&=& {\rm Res}_b <\Psi^{(-1)} \delta \Psi> \wedge \delta\mu \,
d\la \\
&& \hskip -2cm -\, {\rm Res}_b <\Psi^{(-1)} \delta\left({d\la\over d\mu}
{dL\over d\la} \Psi\right)> \wedge \delta\mu \, d\la
\end{eqnarray*}
The first term exactly cancels the term 
${\rm Res}_b K_2$. The second term gives a non--vanishing
contribution 
$${\rm Res}_b \, {\delta \mu_b \over \mu-\mu_b} \wedge \delta\mu \,
d\la$$ 
Indeed let us call $\zeta=(d\la/d\mu)(dL/d\la)\Psi$ which vanishes at
$b=(\la_b,\mu_b)$. Then $\zeta=(\mu - \mu_b)\zeta_1$ hence 
$\delta\zeta=-\delta\mu_b/(\mu -\mu_b)\zeta+\delta\mu \,\zeta_2+\zeta_3$
with $\zeta_3$ regular. The second term does not contribute due to the
antisymmetry of the wedge product and the third term has no residue.
Using eq.(\ref{variation}) we have $<\Psi^{(-1)}{d\la\over d\mu}{dL\over
d\la}\Psi>=1$ yielding the above formula. This contribution is exactly
canceled by the contribution of $K_3$.

We now compute the residues above $\la=\infty$. Recall that we
consider a reduced Hamiltonian system under the action of diagonal
matrices. To fix this action on can normalize the eigenvectors at
$\infty$ so that $\psi_i(Q_j)=\la\delta_{ij}+O(1)$ for $i,j=2, \cdots,
N$. Notice that $L=L_0+O(1/\la)$ where $L_0$ is non--dynamical so 
$\delta L_0=0$, and that $\mu=a_i+O(1/\la)$ around $Q_i$ hence
$\delta L = \delta\mu =O(1/\la)$. Moreover $\Psi^{(-1)}$ vanishes at
$Q_i$ and $d\la$ has a double pole. Altogether $K_1$ and $K_{2}$
are regular at $Q_i$
since $(\delta \Psi)(Q_i)=O(1)$ due to the normalization condition.
Finally $K_{3}$ is also regular since on the sheet $\mu = \mu_{i}(\la)$
one can write $\partial_{\mu} R = \prod_{j\neq i} (\mu_{i} - \mu_{j})$ 
yielding $\delta \log \partial_{\mu} R = O(1/\la)$. All this shows 
that $K$ has no residues above $\la = \infty$.

\proclaim Proposition.
The symplectic form $\omega$ is given by:
\begin{eqnarray}
\omega = 2 \sum_{i=1}^g \delta \lambda_i \wedge \delta \mu_i = 
2 \sum_k {\rm Res}_{\lambda_k}\,{\rm Tr}
\((A_k)_- g_k^{-1}\delta g_k \wedge g_k^{-1}\delta g_k\)  \,d \lambda  \label{symp}
\end{eqnarray}
where $(\lambda_i, \mu_i)$, $i = 1, \cdots g$, are the coordinates of
the points of the dynamical divisor $D$.
\par
\proof
Let us  compute the residues at the poles $\lambda_k$ of $K_{1}$, where 
only $L_k$ contributes. Since locally we have $L = g_k A_k g_k^{-1} $ around
$\la=\la_k$, we may identify the matrix
$\hat \Psi(\la)$ with $g_k$. More precisely  we have
$\widehat{\Psi}(\lambda) = g_k d_k$ and $\widehat{\Psi}^{-1}(\lambda)
= d_k^{-1} g_k^{-1}$ with $d_k$ a diagonal matrix. The residues are
obtained by integrating over small circles surrounding each of the $N$
points $P_{k,i}$ above $\lambda_k$. We can choose these small circles
so that they project on the base $\lambda$ on a single small circle
surrounding $\lambda_k$. Then we get
\begin{eqnarray}
\sum_{i=1}^N {\rm Res}_{P_{k,i}} K_{1} &=& \sum_{i=1}^N {1\over
2i\pi}\oint_{C_{k,i}} <   \Psi^{(-1)}(P_i) \delta L(\lambda)
\wedge \delta \Psi(P_i) >\, d\la
\nonumber \\ &=& {1\over 2i\pi}\oint_{C_k} {\rm Tr} \( 
\widehat{\Psi}^{-1}(\lambda) \delta L(\lambda) \wedge \delta 
\widehat{\Psi}(\lambda) d\la \) \label{completeness}
\end{eqnarray}
where we used the fact that $\hat \Psi^{-1}(\la)$ is equal to the
matrix whose lines are the vectors $\Psi^{(-1)}(P_i) $. The trace has
been reconstructed in eq.(\ref{completeness}) because $\Psi(P_{i})$,
$i=1,\cdots N$, form a basis of eigenvectors. Using the identification
of $\hat \Psi(\la)$ in terms of $g_k$ gives:
\begin{eqnarray}
{\rm Res}_{\lambda_k} K_{1} &=& {\rm Res}_{\la_k}\, {\rm Tr}
\(  d_k^{-1}g_k^{-1}
\; \( \delta g_k (A_k)_- g_k^{-1} -
g_{k}(A_k)_- g_k^{-1} \delta g_k g_k^{-1} \) \wedge \non\\
&& \hskip 3cm
\( \delta g_k d_k + g_k \delta d_k \) \)\,d\la \non\\ &=&
- 2\, {\rm Res}_{\lambda_k}\,  {\rm Tr} \( (A_k)_- g_k^{-1} \delta 
g_k \wedge g_k^{-1} \delta g_k \)\,d\la \non\\ &&\quad +
{\rm Res}_{\lambda_k}\,{\rm Tr} \(g_k^{-1} \delta g_k [ (A_k)_-, 
\delta d_{k} d_{k}^{-1} ]   \)\,d\la
\end{eqnarray}
The last term vanishes because it involves the commutator of two 
diagonal matrices.
To compute the residues of $K_{2}$  at $\la_{k}$ we remark 
that $\delta\mu$ is regular on all sheets above $\la_{k}$. This is 
because due to the form of $L(\la)$, we have $\widehat{\mu} = 
(A_k)_- + {\rm regular}$. Since $(A_k)_-$ characterizes the coadjoint 
orbit and is not dynamical, one has to take $\delta (A_k)_- = 0$. It 
follows immediately that $K_{2}$ is regular. To compute the residue 
of $K_{3}$ we note that if $\partial_{\mu} R$ has a pole of some order $m$
at $P_{k,i}$ and can be written $\partial_{\mu} R = c(\la) /(\la - 
\la_{k})^m$ where $c(\la)$ is regular and non vanishing, we get
$\delta\(\log\, \partial_\mu R\)= \delta \log c(\la)$ which is
regular, since $\delta\la=0$ and $\delta\la_k=0$.  Hence $K_3$ has no
residue again because $\delta\mu$ is regular.

This proposition means that the coordinates $(\la_i,\mu_i)$ of the
point $\nu_i$ of the dynamical divisor are canonical coordinates.
This type of result can be obtained in the r--matrix approach for
specific models like the Toda chain~\cite{Skl}.

\section{Conclusion}

We have shown under quite general but necessary assumptions
that the natural symplectic structure on Lax pairs can be expressed in
terms of algebro--geometric data.
 This result shows the nice interplay between the
analytical and the group--theoretical approaches to integrable
systems. We are able to show that $(\la_i,\mu_i)$ are canonical
coordinates {\em only} using the fact that $L$ parametrizes a
coadjoint orbit, specified by {\em constant} matrices $(A_k)_-$ and
$L_0$. 

\bigskip
\noindent{\bf Acknowledgements.} We thank J. Avan, D. Bernard,
I. Krichever and F. Smirnov for many interesting discussions on the
subject of this paper.

\end{document}